\documentclass[pra,twocolumn]{revtex4-1}
\usepackage{color,float,graphicx,amsmath,bbold}
\begin{document}
\title{Global aspects of polarization optics and coset space geometry}
\author{Arvind}
\email{arvind@iisermohali.ac.in}
\affiliation{Department of Physical Sciences,
Indian
Institute of Science Education \&
Research (IISER) Mohali, Sector 81 SAS Nagar,
Manauli PO 140306 Punjab India.}
\author{ S. Chaturvedi}
\email{subhash@iiserbhopal.ac.in}
\affiliation{Department of Physics, 
Indian
Institute of Science Education \&
Research (IISER) Bhopal, Bhopal Bypass Road,
Bhauri, Bhopal 462066 India}
\author{N. Mukunda}
\email{nmukunda@gmail.com}
\affiliation{
INSA C V Raman Research Professor,
Indian Academy of Sciences,
C V Raman Avenue, Sadashivanagar,
Bangalore 560080 India}
\begin{abstract}
We use group theoretic  ideas and coset space
methods to deal with problems in polarization
optics of a global nature.  These include the
possibility of a globally smooth phase convention
for electric fields for all points on the
Poincar\'{e} sphere, and a similar possibility of
real or complex bases of transverse electric
vectors for all possible propagation directions.
It is shown that these methods help in
understanding some known results in an effective
manner, and in answering new questions as well.
We find that apart from the groups $SU(2)$ and
$SO(3)$ which occur naturally in these problems,
the group $SU(3)$ also plays an important role.
\end{abstract}
\maketitle
\section{Introduction}
\label{intro}
It is well known that the Poincar\'{e}
sphere~\cite{poincare-book} representation of the
pure polarization states of a plane
electromagnetic wave (with fixed frequency and
propagation direction) is intimately related to
the properties  of the two-dimensional unitary
unimodular group $SU(2)$~\cite{born-book}.
Similarly if one considers choices of transverse
real electric field  vectors corresponding to all
possible propagation directions in space, the real
rotation group $SO(3)$ becomes  relevant. In both
cases, questions of a global nature can be
analysed particularly effectively if they are cast
into a group theoretical form.

In this paper such an approach is used to analyse
three such questions. 
The first is the search for
a globally smooth phase convention for transverse
electric field vectors corresponding to all points
on the Poincar\'{e} sphere
\cite{nityananda-pram-79}. This question is shown
to have a simple and elegant resolution when
examined in the framework of the group $SU(2)$.
The second and third questions are concerned with choices of
globally smooth basis states for transverse
electric field vectors as the propagation
direction varies over all points on the sphere of
directions in physical space. The second question
deals with the case of real fields, while the third
question asks whether such a smooth basis is possible for
complex electric fields.
Here it is a well known result that if we limit
ourselves to real fields (corresponding to linear
polarizations), such bases do not exist because of
an obstruction~\cite{nityananda-annal-14}.

The relevant mathematical result is that the
sphere $\mathbb{S}^2$ is not parallelizable, more
informally that it `cannot be combed'. We will see
that relating this question to $SO(3)$ leads to an
immediate and illuminating understanding of this
result. However it has been pointed out recently
that if one goes to the complex domain and allows
electric field vectors in all possible (pure)
polarization states, the obstruction
disappears~\cite{mukunda-josa-14}.  
There exist bases of complex electric fields which
are globally smooth over the sphere of directions.
Here too it turns out that the use of the group
$SU(2)$, and to a limited extent of $SU(3)$ as
well, clarifies the situation
greatly~\cite{mukunda-pra-02}. We develop in full
detail a simple and attractive group theory based
choice of a complex globally smooth basis over the
sphere of directions, by combining the answers to
the first two questions in such a way that the
obstructions seen in them compensate, or
annihilate, one another. We also describe an
important property of the set of all such globally
smooth choices.

The material of this paper is organized as
follows. In Section~\ref{phase_conv} we address the first of the
three questions mentioned above and recast it in
the language of group theory. In particular, we
show that the problem of finding a globally smooth
phase convention over the Poincar\'e sphere is
entirely equivalent to the problem of  choosing
representatives of the coset space $SU(2)/U(1)$ in
a manner that is smooth over all points on the
Poincar\'e sphere. This reformulation of the
original question not only permits us to furnish a
simple proof of the well known impossibility of a
globally smooth phase convention over the
Poincar\'e sphere but, on the positive side,
enables us to provide a parameterization that works
at all but one point (the South pole) of the
Poincar\'e sphere. In
Section~\ref{propagation_direction}, restricting
ourselves to real electric fields, we turn to the
question of  finding a real orthonormal basis in
the two dimensional tangent plane at each point on
the sphere of directions in a smooth manner. Here,
in the same spirit as in Section~\ref{phase_conv}, we show that
this exercise reduces to finding a smooth choice
of representatives, this time for the coset space
$SO(3)/SO(2)$. Again, as before, besides providing
a simple proof of the well known non existence of
such bases, the use of group theoretic language
facilitates construction of bases with the desired
properties at all but one point on the sphere of
directions. Next in
Section~\ref{propagation_direction} we reexamine
this question allowing the electric field to be
complex. In other words we ask the question
whether or not it is possible to come up with a
smooth choice of orthonormal bases in the
complexified tangent spaces at each point on the
sphere of directions. The complexification brings
the group $SU(3)$ into play in a natural way and
we show that by a judicious use of the constructs
developed earlier, one is not only able
to answer the question at hand in the affirmative
but is also able to give an elegant and
essentially unique construction of the desired
bases.  Section~\ref{concl} contains our
concluding remarks and further outlook. 

\section{Globally smooth phase convention over the
Poincar\'{e} sphere-- the obstruction}
\label{phase_conv}
Consider plane electromagnetic waves (with fixed
frequency $\omega$ and propagating along the
positive $z$-axis) in various states of pure
polarization. Denote the two-component complex
electric field vector in the transverse $x-y$
plane, say at $z=0$, by
\begin{equation} {\bf E}=\left(\begin{array}{c}
E_x\\E_y\end{array}\right)= \left(\begin{array}{c}
E_1\\E_2\end{array}\right)\in \mathbb{C}^2.
\end{equation}
The intensity in suitable units is
$I={\bf E}^\dag {\bf E}$,
and for simplicity it will hereafter be set equal
to unity. The polarization state is represented by
a point on the Poincar\'{e} sphere
$\mathbb{S}^2_{\rm pol}$, given in terms of ${\bf
E}$ by
\begin{eqnarray} 
& \hat{\bf n}={\bf
E}^\dag\pmb{\tau} {\bf E}\in \mathbb{S}^2_{\rm
pol}.  
\label{n_pol} 
\end{eqnarray}
Here we follow the polarization optics conventions for
${\pmb\tau}$ matrices: 
\begin{equation}
\tau_1=\left(\begin{array}{cc}1 &0\\0&
-1\end{array}\right),
\tau_2=\left(\begin{array}{cc} 0 &1\\1&
0\end{array}\right),
\tau_3=\left(\begin{array}{cc} 0 &-i\\i&
0\end{array}\right)
\end{equation}
which are related by a
cyclic permutation to the Pauli matrices ${\pmb
\sigma}$ normally used in quantum mechanics:
$\tau_1=\sigma_3, \tau_2=\sigma_1,
\tau_3=\sigma_2$.
The three-vector $\hat{\bf n}$
is real, has unit length, and is unchanged by an
overall phase transformation ${\bf E}\rightarrow
{\bf E}'= e^{i\phi}{\bf E}$. Points on the equator
$n_3=0$ (${\bf E}$ real apart from an overall
phase) correspond to linear polarizations. The two
poles $(0, 0, \pm 1)$ represent circular
polarizations, RCP or LCP, for ${\bf
E}=\frac{1}{\sqrt{2}}\left(\begin{array}{c} 1\\\pm
i\end{array}\right)$ apart, again, from overall
phases.  A physically interesting question is
this: for each $\hat{\bf n}\in \mathbb{S}^2_{\rm
pol}$, is it possible to choose ${\bf E}(\hat{\bf
n})\in \mathbb{C}^2$ such that
\begin{equation}
{\bf E}(\hat{\bf n})^\dag{\pmb \tau}{\bf E}(\hat{\bf n})=\hat{\bf n},
\label{smooth_E}
\end{equation}
with ${\bf E}(\hat{\bf n})$ varying smoothly with
respect to $\hat{\bf n}$ for all $\hat{\bf n}$? We
can recast this question in terms of the group
$SU(2)$ as follows. There is a simple one-to-one
correspondence between $SU(2)$ elements and
normalized two-component complex column vectors
(like ${\bf E}$ above):
\begin{eqnarray} 
&&u\in SU(2)\Leftrightarrow \nonumber \\
&&u=
\left(\begin{array}{cc}\alpha &-\beta^*\\\beta&
\alpha^*\end{array}\right),\,
|\alpha|^2+|\beta|^2=1\Leftrightarrow\nonumber\\
&& {\pmb
\xi}=\left(\begin{array}{c}\alpha\\\beta\end{array}\right)\in
\mathbb{C}^2,\quad {\pmb \xi}^\dag{\pmb
\xi}=1;\nonumber\\ && u({\pmb \xi})=({\pmb \xi},
-i\tau_3{\pmb \xi}^*).
\label{u_si_rel}
\end{eqnarray}
This kind of relationship is special to 
$SU(2)$~\cite{chinn-book,eisenberg-ammath-79}, 
and it shows immediately that, since ${\pmb
\xi}\in \mathbb{S}^3$, as a manifold
\begin{equation}
SU(2)\sim \mathbb{S}^3.
\end{equation}
Moreover, $u({\pmb \xi})$ obeys a `covariance condition'
\begin{equation}
S u({\pmb \xi})=u(S{\pmb \xi}), \quad {\rm any}\  S \in SU(2),
\label{covariance}
\end{equation}
which is very useful.

With $u$ and ${\pmb \xi}$ related as in
Eqn.~(\ref{u_si_rel}), we find that
\begin{eqnarray}
u\tau_1u^{-1}&=&\hat{\bf n}\cdot {\pmb
\tau},\nonumber\\ 
\hat{\bf n}&=&{\pmb
\xi}^\dag{\pmb \tau}{\pmb
\xi}=(|\alpha|^2-|\beta|^2, 2 \Re
(\alpha^*\beta), 2\Im
(\alpha^*\beta));\nonumber\\
u'&=&ue^{i\phi\tau_1}\Leftrightarrow {\pmb
\xi}'=e^{i\phi}{\pmb \xi}.
\label{n_su2}
\end{eqnarray}

The set of elements $\{e^{i\phi\tau_1},
0\le\phi<2\pi\}\subset SU(2)$ defines a $U(1)$
subgroup, leading to the coset space $SU(2)/U(1)$
which can be identified with $\mathbb{S}^2_{\rm
pol}$.  We see that $\hat{\bf n}$ in
Eqn.~(\ref{n_su2}) is
determined by ${\pmb \xi}$ in exactly the same way
as Eqn.~(\ref{n_pol}) determines the polarization state of
${\bf E}$. In the present case we view
Eqn.~(\ref{n_su2})
as defining the projection map $\pi:
SU(2)\rightarrow SU(2)/U(1)=\mathbb{S}^2_{\rm
pol}$:
\begin{eqnarray} u\in SU(2)\rightarrow
\pi(u)&=&\hat{\bf n}={\pmb \xi}^\dag{\pmb \tau}{\pmb
\xi}\in \mathbb{S}^2_{\rm pol},\nonumber\\ 
\pi(ue^{i\phi\tau_1})&=&\pi(u).  
\end{eqnarray}
Conversely, each $\hat{\bf n}\in \mathbb{S}^2_{\rm
pol}$ determines a corresponding coset in $SU(2)$:
\begin{eqnarray}
&&\hat{\bf n}\in \mathbb{S}^2_{\rm
pol}\Leftrightarrow \nonumber \\
&&\mathcal{C}(\hat{\bf
n})=\{u\in SU(2)| \pi(u)=\hat{\bf
n}\}=\pi^{-1}(\hat{\bf n})\subset SU(2).
\end{eqnarray}

The search for a smooth ${\bf E}(\hat{\bf n})$ for
all $\hat{\bf n}\in \mathbb{S}^2_{\rm pol}$
obeying Eqn.~(\ref{smooth_E}) is the same as the search for a
smooth ${\pmb \xi}(\hat{\bf n})$ for all $\hat{\bf
n}$ such that
\begin{equation} {\pmb \xi}(\hat{\bf n})^\dag\
{\pmb \tau}\ {\pmb \xi}(\hat{\bf n})=\hat{\bf n}.
\label{smooth_xi}
\end{equation}
Since by Eqn.~(\ref{u_si_rel})  $u$ and ${\pmb \xi}$ determine
one another, this in turn is the same as the
search for a smooth coset representative
$u(\hat{\bf n})\in \mathcal{C}(\hat{\bf n}) $ for
all $\hat{\bf n}$.

However such globally smooth coset representatives
are not possible. Suppose one such, $u_0(\hat{\bf
n})$ say, did exist.  It would then permit the
expression of any $u\in SU(2)$ in a globally
smooth way as a product of a coset representative
and a subgroup element:
\begin{eqnarray}
&&u\in SU(2),\quad \pi(u)= \hat{\bf n}\in
\mathbb{S}^2_{\rm pol}:\nonumber\\ && u=
u_0(\hat{\bf n})e^{i\phi\tau_1},\quad 0\le
\phi<2\pi.
\label{coset_su2}
\end{eqnarray}
This would imply that as a manifold $SU(2)\simeq
\mathbb{S}^3\simeq \mathbb{S}^2\times
U(1)=\mathbb{S}^2\times \mathbb{S}^1$, which is
easily seen to be false. 
Therefore no
globally smooth $\xi(\hat{\bf n})$ obeying 
Eqn.~(\ref{smooth_xi}), or ${\bf E}(\hat{\bf n})$ obeying Eqn.
(\ref{smooth_E}), can be found.

While it is the case that no smooth
coset representative $u_0(\hat{\bf n})$ for all
$\hat{\bf n}\in \mathbb{S}^2_{\rm pol}$ is
available, we can define such a coset
representative for all $\hat{\bf n}$  except, for
example, at the South pole $(0, 0, -1)$ of
$\mathbb{S}^2_{\rm pol}$. Here is a simple
construction, with $\hat{\bf n}$ parametrised by
spherical polar angles $\theta, \phi$:
\begin{eqnarray}
\hat{\bf n}&=&(\sin \theta \cos \phi,
\sin\theta\sin\phi, \cos\theta)\in
\mathbb{S}^2_{\rm pol},\nonumber  \\
&& \qquad 0\le \theta< \pi,
0\le \phi<2\pi;\nonumber\\ 
u_0(\hat{\bf
n})&=&({\pmb \xi}_0(\hat{\bf n}), -i\tau_3{\pmb
\xi}_0(\hat{\bf n})^*),\nonumber\\ 
{\pmb
\xi}_0(\hat{\bf n})&=&S_0\left(\begin{array}{c}\cos
\theta/2\\
e^{i\phi}\sin\theta/2\end{array}\right),
\nonumber 
\end{eqnarray}
\begin{equation}
S_0=\frac{1}{2}(\mathbb{1}+i\tau_1+i\tau_2+i\tau_3)=
\frac{1}{2}\left(\begin{array}{ll}1+i
&1+i\\ -1+i& 1-i\end{array}\right)\in
SU(2)
\label{polar}
\end{equation}
where the matrix $S_0$ connects the ${\pmb \tau}$
and ${\pmb \sigma}$ matrices: \begin{equation}
\tau_j=S_0\sigma_jS_0^{-1}, j=1, 2, 3.
\end{equation} One can check that
Eqns.(\ref{n_su2}) are
obeyed for all $\hat{\bf n}$ except the South
pole. At the North pole, we do find that these
expressions are all well-defined:
\begin{eqnarray}
&&\theta\rightarrow 0:\qquad {\pmb \xi}_0(\theta,
\phi)\rightarrow
\frac{e^{i\pi/4}}{\sqrt{2}}\left(\begin{array}{c}1\\
i\end{array}\right),\nonumber\\ && u_0(\theta,
\phi)\rightarrow S_0.
\label{polar_limit1}
\end{eqnarray}
On the other hand, at the South pole the limiting
expressions are $\phi$ dependent, hence undefined:
\begin{eqnarray}
&&\theta\rightarrow \pi:\qquad {\pmb \xi}_0(\theta, \phi)\rightarrow \frac{e^{i(\pi/4+\phi)}}{\sqrt{2}}\left(\begin{array}{c}1\\
-i\end{array}\right),\nonumber\\
&&u_0(\theta, \phi)\rightarrow -iS_0\tau_3e^{i\phi\tau_1}.
\label{polar_limit2}
\end{eqnarray}
Nevertheless, as we shall see in the next Section,
this construction is useful in another context.
\section{Bases of electric field vectors for 
all propagation directions}
\label{propagation_direction}
We turn next to the consideration of some global
problems connected with the collection of all
plane waves (with fixed frequency $\omega$) with
all possible propagation directions. We denote the
sphere of propagation directions in physical space
by $\mathbb{S}^2_{\rm dir}$, to be distinguished
from the Poincar\'{e} sphere $\mathbb{S}^2_{\rm
pol}$. For any unit vector $\hat{\bf k}\in
\mathbb{S}^2_{\rm dir}$, the complex
three-component electric field ${\bf E}$ (omitting
the plane wave factor $e^{-i\omega(t-\hat{\bf
k}.{\bf x}/c)}$) obeys the transversality
condition
\begin{equation}
\hat{\bf k}\cdot{\bf E}=0,
\end{equation}
so it is essentially two-dimensional.

To begin with, consider real fields. Then the
vector ${\bf E}$ lies in the two-dimensional plane
in physical space tangent to $\mathbb{S}^2_{\rm
dir}$ at $\hat{\bf k}$:
\begin{equation}
{\bf E}\,\, {\rm real},\quad  \hat{\bf k}\cdot
{\bf E}=0\Leftrightarrow {\bf E}\in T_{\hat{\bf
k}} \mathbb{S}^2_{\rm dir}\subset \mathbb{R}^3.
\end{equation}
We now ask whether it is possible to choose an
orthonormal basis of real vectors ${\bf
E}^{(a)}(\hat{\bf k})$ for each $\hat{\bf k}$
obeying
\begin{eqnarray}
&& \hat{\bf k}\cdot{\bf E}^{(a)}(\hat{\bf
k})=0,\quad {\bf E}^{(a)}(\hat{\bf k})\cdot
{\bf E}^{(b)}(\hat{\bf k})=\delta_{ab},
\nonumber\\ && {\bf E}^{(1)}(\hat{\bf k})\
{}_\wedge\ {\bf E}^{(2)}(\hat{\bf k})=\hat{\bf
k},\qquad a, b=1, 2,
\label{E_conditions}
\end{eqnarray}
in a globally smooth way. We have included here
the condition that $({\bf E}^{(1)}(\hat{\bf k}),
{\bf E}^{(2)}(\hat{\bf k}), \hat{\bf k})$ form a
right handed system; they obviously form an
orthonormal system in three dimensions. (Actually,
it suffices to be able to choose one real vector
${\bf E}^{(1)}(\hat{\bf k})$ for each $\hat{\bf
k}$ obeying
\begin{equation}
\hat{\bf k} \cdot  {\bf E}^{(1)}(\hat{\bf
k})=0,\quad {\bf E}^{(1)}(\hat{\bf k})\cdot
{\bf E}^{(1)}(\hat{\bf k})=1.
\end{equation}
Then if we define
\begin{equation}
{\bf E}^{(2)}(\hat{\bf k})=\hat{\bf k}\ {}_\wedge\
{\bf E}^{(1)}(\hat{\bf k}),
\end{equation}
all of Eqns.~(\ref{E_conditions}) are obeyed). If such a choice
were possible, for each $\hat{\bf k}\in
\mathbb{S}^2_{\rm dir}$  we could define an
element $R(\hat{\bf k})\in SO(3)$, the proper real
orthogonal rotation group in three dimensions,
carrying ${\bf e}_3$ to $\hat{\bf k}$:
\begin{eqnarray}
&&\hat{\bf k}\in \mathbb{S}^2_{\rm dir}\rightarrow
R(\hat{\bf k})\in SO(3):\nonumber\\ && R(\hat{\bf
k})({\bf e}_1, {\bf e}_2, {\bf e}_3)=( {\bf
E}^{(1)}(\hat{\bf k}),  {\bf E}^{(2)}(\hat{\bf
k}), \hat{\bf k}).
\end{eqnarray}
Here ${\bf e}_j, j=1,2,3$ are the unit vectors
along the three Cartesian coordinate axes in
physical space. Therefore the columns of the
matrix $R(\hat{\bf k})$ are:
\begin{eqnarray} &R_{j1}(\hat{\bf k})= {\bf
E}_j^{(1)}(\hat{\bf k}), \quad R_{j2}(\hat{\bf
k})= {\bf E}_j^{(2)}(\hat{\bf k}), \quad
R_{j3}(\hat{\bf k})= \hat{k}_j,\nonumber \\
&j=1,2,3.
\label{column_matrix}
\end{eqnarray}
The elements of $SO(3)$ leaving ${\bf e}_3$
invariant are rotations in the $x—y$ plane,
forming an $SO(2)$ subgroup of $SO(3)$:
\begin{eqnarray}
&&SO(2)=\nonumber \\
&&\left\{R_3(\phi)=\left(\begin{array}{lll}\cos\phi&
-\sin\phi & 0\\ \sin\phi& \cos \phi & 0\\ 0& 0&
1\end{array}\right)|0\le \phi<2\pi\right\}\subset
SO(3),\nonumber\\ 
&& \qquad \qquad R_3(\phi){\bf e}_3={\bf e}_3.
\end{eqnarray}
Similar to the situation with $SU(2)$, here too we
have the coset space identification
\begin{equation}
SO(3)/SO(2)\simeq \mathbb{S}^2_{\rm dir}.
\end{equation}
Therefore if a choice of $R(\hat{\bf k})$ smooth
over all $\hat{\bf k}\in S^2_{\rm dir}$ did exist,
it would be a coset representative and we could
use it to express every $R\in SO(3)$ smoothly as a
product,
\begin{eqnarray}
&& R\in SO(3), \quad R{\bf e}_3=\hat{\bf k}\in
S^2_{\rm dir}:\nonumber\\ && R=R(\hat{\bf
k})R_3(\phi),\quad 0\le \phi< 2\pi,
\end{eqnarray}
similar to Eqn.~(\ref{coset_su2}). However this would mean
that $SO(3)$ has the global structure of
$\mathbb{S}^2\times SO(2)$. This conflicts with
the fact that it is $\mathbb{S}^3/\sim$, where
$\sim$ is the identification of antipodal points
on $\mathbb{S}^3$. Therefore globally smooth
choices of $R(\hat{\bf k})\in SO(3)$ obeying 
Eqn.~(\ref{column_matrix}) are not possible.

As mentioned earlier, this is a well-known
result~\cite{nakahara-book}. Another even more elementary analytic proof
is given later.

Now we extend this analysis by considering complex
electric field vectors. This means that at each
$\hat{\bf k}\in \mathbb{S}^2_{\rm dir}$, we go
from the real two-dimensional tangent plane
$T_{\hat{\bf k}}\mathbb{S}^2_{\rm
dir}\subset\mathbb{R}^3$ to its complexification
which is no longer contained in $\mathbb{R}^3$:

\begin{eqnarray}
&&\hat{\bf k}\in \mathbb{S}^2_{\rm dir}:\quad
T_{\hat{\bf k}}\mathbb{S}^2_{\rm dir}\rightarrow
(T_{\hat{\bf k}}\mathbb{S}^2_{\rm dir})^c=
\nonumber\\ && \{ {\bf E}=\ 
\mbox{complex
three-dimensional vector}\ \vert \hat{\bf k}\cdot
{\bf E}=0\}. 
\nonumber \\
\label{complex_E}
\end{eqnarray}
It now turns out that it is possible to find (in
infinitely many ways) orthonormal bases for these
complexified tangent spaces, which are globally
smooth with respect to $\hat{\bf
k}$~\cite{nityananda-annal-14}. The
complexification in Eqn.~(\ref{complex_E}) suggests that it is
useful to extend the groups $SU(2)$ and $SO(3)$ so
far used to $SU(3)$, the group of complex unitary
unimodular matrices in three dimensions. This
contains both $SU(2)$ and $SO(3)$ as subgroups. To
begin with, let us write $\mathcal{A}$ for a
general matrix in the unitary group $U(3)$. If we
demand that the third column $\mathcal{A}_{j3}$ be
the components of a chosen $\hat{\bf k} \in
\mathbb{S}^2_{\rm dir}$, unitarity of
$\mathcal{A}$ guarantees that the first two
columns of $\mathcal{A}$ form an orthonormal basis
for $(T_{\hat{\bf k}}\mathbb{S}^2_{\rm dir})^c$:
\begin{eqnarray}
&& \mathcal{A}_{ja}=E_j^{(a)}, \quad
\mathcal{A}_{j3}=\hat{k}_j:\nonumber\\ &&
\mathcal{A}^\dag \mathcal{A}=\mathbb{1}_{3\times
3}\Leftrightarrow
E_j^{(a)^*}E_j^{(b)}=\delta_{ab},\,
\hat{k}_jE_j^{(a)}=0, \, a, b=1, 2. \nonumber \\
\label{column_basis}
\end{eqnarray}
We then find easily that $\mathcal{A}$ has a rather simple form:
\begin{eqnarray}
\mathcal{A}&=&R
\left(\begin{array}{cc} 
u & \begin{array}{c} 0 \\ 0  \end{array} \\
\begin{array}{cc} 0 & 0 \end{array} & 1
\end{array}\right),
\quad R\in SO(3),
u\in U(2),
\nonumber\\ 
R_{j3}&=&\hat{k}_j.
\label{structure}
\end{eqnarray}
As in Eq.~(\ref{E_conditions}) let us now add the righthandedness condition in the sense
\begin{equation}
\varepsilon_{jkl}\mathcal{A}_{k1}\mathcal{A}_{l2}=\mathcal{A}_{j3}=\hat{k}_j.
\end{equation}
This implies
\begin{equation}
\varepsilon_{jkl}\mathcal{A}_{k1}\mathcal{A}_{l2}\mathcal{A}_{j3}=\det
\mathcal{A}=1,
\end{equation}
that is, $\mathcal{A}\in SU(3)$. Then we find that
in the structure~(\ref{structure}) for $\mathcal{A}$ we must
have $u\in SU(2)$. This will hereafter be assumed.
Such matrices $\mathcal{A}$ form a subset, not a
subgroup, in $SU(3)$. The breakup~(\ref{structure}) of
$\mathcal{A}$ into two factors is however not
unique since there are shared elements:
\begin{equation}
R_3(\phi)= 
\left(\begin{array}{cc} 
e^{-i\phi\tau_3}
 & \begin{array}{c} 0 \\ 0  \end{array} \\
\begin{array}{cc} 0 & 0 \end{array} & 1
\end{array}\right).
\end{equation}
Let us parametrise $\hat{\bf k}\in \mathbb{S}^2_{\rm dir}$ 
in the same way as $\hat{\bf n}\in \mathbb{S}^2_{\rm pol}$ 
in Eqn.~(\ref{polar}):
\begin{eqnarray}
\hat{\bf k}&=&(\sin\theta \cos \phi, \sin \theta\sin \phi, \cos \theta)\in \mathbb{S}^2_{\rm dir},
\nonumber \\
&& 0\le \theta\le \pi, 0\le \phi<2\pi.
\end{eqnarray}
Momentarily avoiding the South pole $\theta=\pi$, let us define $R_0(\theta, \phi)\in SO(3)$ as
\begin{eqnarray}
R_0(\theta, \phi)&=&R_3(\phi)R_2(\theta)R_3(\phi)^{-1},\nonumber\\
R_2(\theta)&=& \left(\begin{array}{ccc}\cos \theta &0 &\sin \theta\\ 0&1& 0\\-\sin\theta& 0& \cos \theta\end{array}\right)\in SO(3),
\nonumber\\
R_0(\theta, \phi){\bf e}_3&=&\hat{\bf k}.
\end{eqnarray}
At the North pole $R_0(\theta, \phi)$ is obviously
well-defined: $R_0(0, \phi)=\mathbb{I}$. However
as we approach the South pole we find a
multivaluedness:
\begin{eqnarray}
\theta\rightarrow \pi: R_0(\theta,
\phi)&\rightarrow& \left(\begin{array}{ccc}-\cos
2\phi &-\sin 2\phi& 0\\ -\sin2\phi&\cos 2\phi &0\\
0& 0& -1\end{array}\right)\nonumber\\ 
&&=\left(
\begin{array}{cc} 
-\tau_1
e^{2i\phi\tau_3}
& \begin{array}{c} 0 \\ 0  \end{array} \\
\begin{array}{cc} 0 & 0 \end{array} & 1
\end{array}\right).
\end{eqnarray}
Let us then write
\begin{eqnarray}
&& \mathcal{A}(\theta, \phi)= R_0(\theta, \phi)
\left(\begin{array}{cc} 
u(\theta,\phi)
& \begin{array}{c} 0 \\ 0  \end{array} \\
\begin{array}{cc} 0 & 0 \end{array} & 1
\end{array}\right)
\in SU(3),\nonumber\\ &&
u(\theta, \phi)\in SU(2), \qquad
\mathcal{A}(\theta, \phi){\bf e}_3=\hat{\bf k}.
\label{su3}
\end{eqnarray}
The conditions for $\mathcal{A}(\theta, \phi)$ to
be well-defined and globally smooth over
$\mathbb{S}^2_{\rm dir}$ are, apart from smooth
dependences on $\theta$ and $\phi$:
\begin{eqnarray}
0<\theta<\pi& \ : \ & \mathcal{A}(\theta, \phi+2\pi)=\mathcal{A}(\theta, \phi);\nonumber\\
 \theta\rightarrow 0&\ :\ &\mathcal{A}(\theta, \phi)\rightarrow 
\left(\begin{array}{cc} 
u(0,\phi)
& \begin{array}{c} 0 \\ 0  \end{array} \\
\begin{array}{cc} 0 & 0 \end{array} & 1
\end{array}\right)
\nonumber \\&&=\phi{\rm -independent};\nonumber\\
\theta\rightarrow \pi&\ : \ & \mathcal{A}(\theta, \phi)\rightarrow 
\left(\begin{array}{cc} 
-\tau_1 e^{2i\phi\tau_3}u(\pi,\phi)
& \begin{array}{c} 0 \\ 0  \end{array} \\
\begin{array}{cc} 0 & 0 \end{array} & 1
\end{array}\right)
\nonumber\\&&=\phi {\rm -independent}.
\end{eqnarray}
These translate into conditions on 
$u(\theta, \phi)$:
\begin{eqnarray}
0<\theta<\pi&\ :\ &u(\theta, \phi+2\pi)=u(\theta, \phi);\nonumber\\
\theta\rightarrow 0&\ :\ &u(0, \phi)=u_0\in SU(2);\nonumber\\
\theta\rightarrow \pi&\ :\ &u(\pi, \phi)=e^{-2i\phi\tau_3}u_\pi, u_\pi\in SU(2).
\label{u_conditions}
\end{eqnarray}

We will develop an expression for $u(\theta,
\phi)$ using the results of Section II, but let us
now show by a simple argument the existence of an
obstruction in the real domain. Such a possibility
would correspond to $u(\theta,
\phi)=e^{-i\alpha(\theta, \phi)\tau_3}\in
SO(2)\subset SU(2)$. Then the
conditions~(\ref{u_conditions})
require that $\alpha(\theta, \phi)$ obey, apart
from smoothness in $\theta$ and $\phi$:
\begin{eqnarray}
 0<\theta<\pi&:& \alpha(\theta, \phi+2\pi)-
\alpha(\theta, \phi)=2n\pi, \nonumber \\
&&n\ \mbox{integer independent of}\ \theta;\nonumber\\
 \theta\rightarrow 0&:& \alpha(0, \phi)=a;\nonumber\\
 \theta\rightarrow \pi&:& \alpha(\pi, \phi)=2\phi+b;\quad a, b\ {\rm constants}.
\end{eqnarray}
The $\theta$-independence of $n$ follows from
smoothness (continuity) in $\theta$. However these
conditions on $\alpha(\theta, \phi)$ are mutually
inconsistent:
\begin{eqnarray}
&& \alpha(0, \phi)=a\Rightarrow n=0;\nonumber\\
&& \alpha(\pi, \phi)=2\phi+b\Rightarrow n=2.
\end{eqnarray}
This is therefore a simple analytic proof of the
nonparallelizability of $T\mathbb{S}^2_{\rm dir}$,
as mentioned earlier.

We now return to the problem of constructing a
group element $u(\theta, \phi)\in SU(2)$ obeying
the conditions~(\ref{u_conditions}). For this we exploit the
construction of $u_0(\theta, \phi)$ in
Eqn.~(\ref{polar}),
with behaviours as $\theta\rightarrow 0, \pi$ as
given in Eqns.~(\ref{polar_limit1},\ref{polar_limit2}). Apart from a scale
change in $\phi$, these are qualitatively similar
to the desired behaviours of $u(\theta, \phi)$. If
we note that
\begin{eqnarray}
&& \tau_3e^{i\phi\tau_1}=e^{-i\phi\tau_1}\tau_3,\nonumber\\
&& e^{-i\phi\tau_1}= e^{-i\frac{\pi}{4}\tau_2} e^{-i\phi\tau_3} e^{i\frac{\pi}{4}\tau_2},
\end{eqnarray}
we can see that a possible solution to our problem
is \begin{equation}u(\theta, \phi)=
e^{i\frac{\pi}{4}\tau_2}S_0^{-1}u_0(\theta, 2\phi)
e^{-i\frac{\pi}{4}\tau_2}.
\label{possible_solution}
\end{equation}
As long as $\theta$ remains in the range $0\le
\theta\le \pi$, for $0<\theta<\pi$ the $2\pi$
periodicity of $u(\theta, \phi)$ in $\phi$ is
satisfied. At the poles we find:
\begin{eqnarray}
&& u(\theta, \phi)\stackrel{\theta\rightarrow 0}{-\!\!\!-\!\!\!-\!\!\!\rightarrow} \mathbb{1},\quad {\rm i.e.}, u_0=\mathbb{1};\nonumber\\
&& u(\theta, \phi) \stackrel{\theta\rightarrow
\pi}{-\!\!\!-\!\!\!-\!\!\!\rightarrow}e^{-2i\phi\tau_3}\cdot
  \tau_2\tau_3,\quad {\rm i.e.}, u_\pi=i\tau_1.
\end{eqnarray}
Thus $u(\theta, \phi)$ is indeed a smooth function
of $(\theta, \phi)\in \mathbb{S}^2_{\rm dir}$ at
all points except the South pole, with the desired
multivaluedness at that pole. The scale change
$\phi\rightarrow 2\phi$ involved in
Eqn.~(\ref{possible_solution})
does not cause any difficulties.
We now use Eqn.~(\ref{possible_solution}) in Eqn.~(\ref{su3}) to get a
globally smooth $SU(3)$ element
$\mathcal{A}_0(\theta, \phi)$ obeying
Eqns.~(\ref{column_basis}).
After some simplifications using the covariance
condition~(\ref{covariance}) we arrive at the expressions:
\begin{eqnarray}
&& u_0(\theta, \phi)=S_0 e^{-i\frac{\phi}{2}\tau_1}
e^{-i\frac{\theta}{2}\tau_3}e^{i\frac{\phi}{2}\tau_1};
\nonumber\\
&& u(\theta, \phi)= e^{-i\phi\tau_3} 
e^{i\frac{\theta}{2}\tau_1} e^{i\phi\tau_3};
\nonumber\\
&& \mathcal{A}_0(\theta, \phi)= R_0(\theta, \phi)
\left(\begin{array}{cc} 
u(\theta, \phi)
& \begin{array}{c} 0 \\ 0  \end{array} \\
\begin{array}{cc} 0 & 0 \end{array} & 1
\end{array}\right) .
\end{eqnarray}
At the poles, $\mathcal{A}_0(\theta, \phi)$ smoothly approaches the unambiguous limiting values
\begin{equation}
\mathcal{A}_0(0, \phi)=\mathbb{1},\, \mathcal{A}_0(\pi, \phi)={\rm diag}(-i, -i, -1).
\end{equation}
Thus according to Eqn.~(\ref{column_basis}), at the poles we have
the orthonormal bases for the complexified tangent
spaces given by
\begin{equation}
{\bf E}^{(a)}(0, \phi)={\bf e}_a, {\bf E}^{(a)}(\pi, \phi)=-i {\bf e}_a,\quad a=1, 2.
\end{equation}
All of these correspond to linear polarizations. On the other hand, at points along the equator $\theta=\pi/2$, we have 
($S=\sin \phi, C=\cos \phi, 0\le \phi<2\pi$):
\begin{eqnarray}
 R_0\left(\frac{\pi}{2},  \phi\right) 
&=&\left(\begin{array}{ccc}S^2 &-SC &
C\\-SC &C^2&S\\-C& -S& 0\end{array}\right);\nonumber\\
u\left(\frac{\pi}{2},  \phi\right) &=&
\frac{1}{\sqrt{2}}\left(\begin{array}{ll}1+i(C^2-S^2) &2iSC \\2iSC &1-i(C^2-S^2)\end{array}\right);
\nonumber\\
\mathcal{A}_0\left(\frac{\pi}{2},\phi\right) &=&
e^{-i\pi/4}\left(\begin{array}{ccc}S^2 &-SC
&e^{i\pi/4}C\\-SC &C^2&e^{i\pi/4}S\\-iC& -iS& 0
\end{array}\right).
\end{eqnarray}
The complex orthonormal basis vectors for the
complexified tangent space to $S^2_{\rm dir}$ at
the equatorial point $(\pi/2, \phi)$ are thus
\begin{eqnarray}
{\bf E}^{(1)}(\pi/2, \phi)&=&e^{-i\pi/4}(S^2, -SC, -iC),\nonumber\\
{\bf E}^{(2)}(\pi/2, \phi)&=&e^{-i\pi/4}(-SC,
C^2, -iS),\nonumber \\
S&=&\sin\phi, C=\cos\phi.
\end{eqnarray}
Each of these corresponds to a pure polarization
state that changes smoothly as $\phi$ varies. For
$\phi=0, \pi/2, \pi, 3\pi/2$ we have linear
polarizations; for $\phi=\pi/4, 3\pi/4, 5\pi/4,
7\pi/4$ we have circular polarizations (of
opposite senses for ${\bf E}^{(1)}$ and ${\bf
E}^{(2)}$); and for all other $\phi$ we have
elliptic polarizations. Since we have a
continuously changing propagation vector $\hat{\bf
k}=(\cos\phi, \sin \phi, 0)$, it is not meaningful
to display all these features on any common or
fixed Poincar\'{e} sphere.

In a similar manner, the behaviours of
$\mathcal{A}_0(\theta, \phi), {\bf
E}^{(a)}(\theta, \phi)$ at other points on
$S^2_{\rm dir}$ can be examined, but we forego the
details.

From the manner in which the $SU(2)$ elements
$u(\theta, \phi)$ and the $SU(3)$ elements
$\mathcal{A}_0(\theta, \phi)$ have been
constructed, it would seem that the solution given
above to the problem of defining globally smooth
polarization bases over all of $S^2_{\rm dir}$ is
in some sense both minimal and natural. The most
general globally smooth $\mathcal{A}(\theta,
\phi)$ is clearly of the form
\begin{equation}
\mathcal{A}(\theta, \phi)= \mathcal{A}_0(\theta, \phi) 
\left(\begin{array}{cc} 
u'(\theta, \phi) 
& \begin{array}{c} 0 \\ 0  \end{array} \\
\begin{array}{cc} 0 & 0 \end{array} & 1
\end{array}\right) ,
\end{equation}
where $u'(\theta, \phi)$ is any globally smooth
map $\mathbb{S}^2\rightarrow SU(2)\simeq
\mathbb{S}^3$. Any number of examples of such
$u'(\theta, \phi)$ are easily constructed; for
instance
\begin{equation}
u'(\theta, \phi)=\exp(i\sin\theta\ {\bf a}
(\theta, \cos \phi, \sin \phi)\cdot {\pmb
\tau})
\end{equation}
for any real ${\bf a} (\theta, \cos \phi, \sin
\phi)$ smooth in $\theta$ and polynomial in
$\cos\phi$ and $\sin \phi$ is acceptable.

It is a result of homotopy theory that any two
globally smooth choices of $ u'(\theta, \phi)$ can
be continuously deformed into one another; the
reason is that the homotopy group
$\pi_2(\mathbb{S}^3)$ is
trivial~\cite{nakahara-book}. Thus we
see that while there are infinitely many choices
of globally smooth complex orthonormal bases
$\{{\bf E}^{(a)}(\hat{\bf k})\}$  for electric
field vectors for plane waves for all $\hat{\bf
k}\in \mathbb{S}^2_{\rm dir}$, any two choices can
be smoothly deformed into one another. In this
precise sense, the solution given above using
$\mathcal{A}_0(\theta, \phi)$ is essentially
unique; any other solution is in the same homotopy
class as this one, indeed there is only one
homotopy class.
\section{Concluding remarks}
\label{concl}
We have looked at three physically important and
interesting problems in classical polarization
optics, all of which have a global character. 
Our principal results, 
can be summarised as follows:
\begin{enumerate}
\item[{[A]}]  The first problem regarding the
Poincar\'e sphere was treated in~\cite{nityananda-pram-79}.  It was
shown there that  there is no globally smooth way
of defining electric field vectors covering the
entire Poincar\'e sphere.  Of course, we cannot
change this result, but we give a much simpler
proof than that in~\cite{nityananda-pram-79}, using SU(2) coset
space properties. We also construct an almost
globally smooth coset representative, good
everywhere except at South pole, which is used
later for a different problem.
\item[{[B]}] As to the second problem, here again
we have the well known result : no globally smooth
choice of real tangent vectors is possible over the
entire sphere of directions.  Here we give a
simple proof using $SO(3)$ coset space, as well as
a simple analytic proof. Admittedly, the negative
result remains unchanged but we go a step further in
that we explicitly construct an almost globally
smooth  $SO(3)$ coset representative, problem only
at South pole, for use in the next problem.
\item[{[C]}]  If in B above one goes from real to
complex tangent vectors, it was shown
in~\cite{nityananda-annal-14}
that  globally smooth bases do exist.  We give an
explicit construction of such a basis by combining
the almost globally smooth expressions in A and B
, using the properties of $SU(3)$, to get a truly
globally smooth solution for problem C. The South
pole problems in A and B can be made to annihilate
one another! The novelty is in use of SU(3), and
the last result is that any two global choices can
be smoothly deformed into one another; in that
sense our group theory based solution is
essentially unique.
\end{enumerate}
The fact that the three-dimensional groups $SU(2)$
and $SO(3)$ play important roles in these problems
comes as no surprise. On the other hand, the use
of the eight-dimensional group $SU(3)$, in a
limited way, in answering the third question is
particularly interesting. It is likely that such
approaches will be found useful in other problems
in the polarization optics context as well.

It is likely that such approaches will be found
useful in other problems in polarization optics
such as in the Poincare' sphere based descriptions
and its generalizations to situations which
involve orbital angular momentum and higher order
effects~\cite{padgett-optlett-99,agarwal-josa-99,milione-prl-11,dennis-royal-17}.
\begin{acknowledgments} Arvind acknowledges
funding from DST India under Grant No.
EMR/2014/000297.  NM thanks the Indian National
Science Academy for enabling this work through the
INSA C V Raman Research Professorship.
\end{acknowledgments}
%
\end{document}